\documentstyle[12pt,aaspp]{article}
\input{psfig}
\def\gax    {${_>\atop^{\sim}}$}
\def\aox    {$\alpha_{ox}$}
\def\etal   {{\it et al.}}

\begin{document}

\title{X-ray Absorption Towards The Einstein Ring Source PKS1830-211}
\author{Smita Mathur$^1$}
\affil{Harvard-Smithsonian Center for Astrophysics,
60 Garden St., Cambridge, MA 02138}
\author{Sunita Nair$^2$}
\affil{University of Manchester, N.R.A.L. Jodrell Bank, Macclesfield,
Cheshire SK11 9DL, U.K.}
\footnotetext[1]{smita@cfa.harvard.edu}
\footnotetext[2]{sunita@jb.man.ac.uk}
\setcounter{footnote}{1}

\received{}
\accepted{}
\lefthead{Mathur \& Nair}
\righthead{X-ray Absorption in PKS1830-211}

\section*{Abstract}

PKS1830-211 is an unusually radio$-$loud gravitationally lensed quasar.
In the radio, the
system appears as two compact, dominant features surrounded by relatively
extended radio emission which forms an Einstein Ring. As the line of sight
to it passes close to our Galactic center, PKS1830-211 has not been
detected in wave-bands other than the radio and X-ray so far. Here we
present X-ray data of PKS1830-211 observed with ROSAT PSPC. The X-ray
spectrum shows that absorption in excess of the Galactic contribution is highly
likely, which at
the redshift of the lensing galaxy ($z_l=0.886$) corresponds to
N$_H=3.5^{+0.6}_{-0.5}\times 10^{22}$ atoms cm$^{-2}$. The
effective optical extinction is large, A$_V$(observed)\gax5.8.
When corrected
for this additional extinction, the two point optical to X-ray slope
\aox\ of PKS1830-211 lies just
within the observed range of quasars. It is argued here that both compact
images must be covered by  the X-ray absorber(s)
 which we infer to be the lensing galaxy(galaxies). The dust to gas ratio
along the line of sight within the lensing galaxy is likely to be
somewhat larger than for our Galaxy.

\newpage
\section{Introduction}

PKS1830-211 (Rao \& Subrahmanyan 1988, Subrahmanyan et al. 1990, Jauncey
et al. 1991) is one of the strongest radio sources in the sky at centimeter
wavelengths. It has been modeled as a gravitationally lensed quasar consisting
of a core, a knot and a jet on the scale of a few hundred
milliarcseconds
(Subrahmanyan et al. 1990 (SNRS90), Nair, Narasimha \& Rao 1993 (NNR93),
Kochanek \& Narayan 1992 (KN92)), which is largely doubly$-$imaged
by a foreground lens galaxy into a system dominated by two compact
features an arcsecond apart. These compact structures constitute the
NE (`north east')
and SW (`south west') images. Each image is resolvable on the
scale of $\sim 100-200$ mas into a knot and a core image, with
contributions from
the relatively flat$-$spectrum core tending to dominate at higher
frequencies of
observation. The steeper spectrum extended emission from the jet is, in part,
quadruply$-$imaged to form a radio Einstein Ring that surrounds the
compact images. As PKS1830-211 is seen close to the Galactic Center,
there has been little success with optical or infrared identification
(Djorgovski et al. 1992). Wiklind \& Combes (1996) (WC96)  report the
detection of a galaxy against
the background of the source in molecular absorption line studies,
providing a lens redshift, $z_l$, of 0.886. According to WC96,
the absorbing gas covers only the SW compact feature.
Such an occurrence would be favoured if the SW image lies closer to
the center of the absorbing galaxy than the NE image.
This is consistent with the modeling in NNR93 and the observed order
of flux density variations in the compact features, as detected by
van Ommen et al. (1995).

Recently, the report of
a possible intervening galaxy at a second redshift ($z=0.19$), detected
in the neutral hydrogen studies of
Lovell et al. (1996), suggests that PKS1830-211 could be a more complex system
than previously thought. The second intervenor, it is argued,
principally intercepts
the NE compact feature rather than the SW one.
However, the observed angular scale of $1^{\prime \prime}$
for the system is typical of a low mass lens like an isolated  spiral
galaxy, and lens modeling of this scale of structure is curiously in
accord with this picture
(NNR93, KN92).
The possibility of multiple lenses being
involved, however, is not excluded.
A time delay for correlated flux
density variations between the two dominant images of the core in the
system has
been reported by van Ommen et al. (1995), with a value of $44\pm9$ days.
The source redshift is unknown as yet; knowledge of this quantity is a
vital clue that will enable us to use this system to estimate or
constrain parameters
of cosmological models, such as Hubble's Constant (Refsdal 1964, 1966).

The single-lens model
of NNR93, coupled with the known lens redshift and time delay between
the core images, predicts a source redshift $z_s$ of around
1.5$-$2.4 (with H$_o =50$ km/s/Mpc, q$_o=0.5$). This model also
suggests that, taken together,
the pair of core images is brighter than their source by a factor of
about $8-10$.

Molecular absorption in the  lensing galaxy (WC96) corresponds to an
equivalent H$_2$ column density of $3\times 10^{22}$ ~cm$^{-2}$. An X-ray
observation of PKS1830-211 would be particularly interesting because (1)~
hard X-rays can penetrate through such a column density making X-ray
detection of the quasar possible (2)~the large column density can be
directly measured, giving us valuable information about the lensing galaxy.
 Since X-ray absorption is
relatively insensitive to moderate ionization and depletion (Morrison
\& McCammon 1983), X-ray spectra give the {\it total} column density of
the absorber. We searched X-ray databases to look for possible
detection of PKS1830-211 with this aim in mind. Even though the source
is located  close to the Galactic center, it is 5.7 degrees out of the
Galactic plane. The Galactic extinction is moderate
with A$_V = 2.7$ magnitudes (SNRS90) making X-ray detection a viable
possibility.
We found that the source was indeed observed by ROSAT (Trumper 1983)
Position Sensitive Proportional Counter (PSPC,
Pfefferman et al. 1987), and was also detected in the ROSAT All Sky
Survey (Brinkmann, Siebert \& Boller 1994).
In this paper we report the pointed PSPC observations, argue that the
X-ray source
is indeed the quasar and discuss the implications of the observations.

\section{{\it ROSAT} Observations and Data Analysis}

 PKS1830-211 was observed with the {\it ROSAT} PSPC on  15 September 1993
for a total livetime of 17,661 sec. over a real time span of 128,320
sec ($\sim$1.5 days).  We
retrieved the data from the HEASARC\footnote {High Energy Astrophysics
Science Archive Research Center is a service of the Laboratory for
High Energy Astrophysics (LHEA) at NASA/GSFC.} database and analyzed it
using the PROS\footnote {Post-Reduction Offline Software}
package in IRAF. The source was clearly detected.
The source centroid position was 18 33 39.4, -21 03 40.7 (J2000) only
7.5$^{\prime\prime}$ off the radio source coordinates of 18 33 39.94, -21 03
40.40 (van Ommen \etal~1995).  This is well
within the pointing accuracy of $\sim$25$^{\prime\prime}$ of ROSAT PSPC.
The source counts were extracted
from within a 2$^{\prime}$  radius circle centered on the source
centroid. The small background was estimated from an annulus centered on the
source and inner and outer radii of 3$^{\prime}$ and 5$^{\prime}$
 respectively. The total net counts were  1174$\pm$40 yielding
a count rate of $0.07\pm0.003$ s$^{-1}$.
We also extracted the light curve of the source. It was  essentially
flat and clearly without any flares.
We used the Kolmogorov-Smirnov and Cramer-von Mises one-sample
goodness-of-fit tests (using IRAF task VARTST) to conclude that the
source is not variable.

 It was immediately apparent from the pulse height distribution that
the X-ray source contained virtually no photons below $\sim0.5$ keV
(Figure 1).
This obviously suggests large low-energy photoelectric absorption. We
made a number of spectral fits to the pulse height data, the results
of which are given in Table 1.
All the spectral fits were made to the $3-34$ PHA
channels as extracted by the standard PROS analysis. Channels 1 and 2
were ignored since they are inadequately calibrated. Channels with few
counts were combined to ensure at least 10 counts per channel, so that
$\chi^2$ statistics could be applied. A standard flat systematic error
of 1\% was added.  The
response matrix released in January 1993 was employed. The errors in
Table 1 represent the 90\% confidence interval. All the spectral
analysis was done using XSPEC, the X-ray spectral analysis software.

 Our first fit was a simple power-law with absorption by cold material at
zero redshift. Both the power-law slope and the absorbing column
density were free to vary. The resulting fit was good
($\chi^2_{\nu}=0.76$), but the best fit absorbing column density was
$\sim$3 times larger than the Galactic value of N$_H=2.63\times
10^{21}$ atoms cm$^{-2}$ (Stark \etal~ 1992). If the fit is made with
Galactic column
density then the resulting value of power-law energy index is $\alpha_x=-1.2$
(Table 1). This is an unusually flat spectrum even for radio-loud
quasars (e.g. Schartel \etal~1996, Shastri \etal~1993). Even if the
Galactic column density is larger, corresponding to the optical
extinction of A$_V = 2.7$ magnitudes (N$_H=4.32\times 10^{21}$ atoms
cm$^{-2}$), the power-law energy index is still unusual ($\alpha_x <
-0.2$). Excess absorption
is thus highly likely and an obvious site for this is within the
lensing galaxy. We thus made another power-law fit with a $z=0$
absorption component fixed at the Galactic value plus another absorption
component, which was allowed to be free, at the redshift of $z_l=0.886$
 of the lensing galaxy. The column density at $z_l=0.886$ is
N$_H=1.7^{+2.9}_{-1.5}\times 10^{22}$ atoms cm$^{-2}$. The power-law
slope $\alpha_x=0.02$ is still extremely flat. If $\alpha_x$ is fixed
to a mean value of 1.2 (Schartel \etal~1996) for radio-loud quasars,
the column density N$_H=3.5^{+0.6}_{-0.5}\times 10^{22}$ atoms
cm$^{-2}$.

 It should be noted, however, that the HI survey by Stark et
al. (1992) has a spetial resolution of 2 degrees. It may well be
possible that small scale structures within this beam exist and that
the Galactic column density towards PKS1830-211 is actually much
higher than the value quoted above. In this case the X-ray
observations presented here imply that the Galactic column density is about
$10^{22}$ cm$^{-2}$ (for $\alpha_x$=1.2, Table 1). This is unlikely
since the corresponding optical extinction in the Galaxy would be much
larger (see \S 3.2) than that reported by SNRS90, and confirmed by
Djorgovski et al. 1992.

 The redshift of the absorber is not
constrained by the X-ray observations. It could be anywhere along the
line of sight, from z=0 to the source redshift. If it is at the
redshift of z=0.19, associated with the absorber of Lovell et
al. (1996), the column density would be $\sim 1.1 \times 10^{22}$
cm$^{-2}$ (for $\alpha_x=1.2$, Table 1.). The  HI column density
 in the z=0.19 absorber, assuming a spin temperature of about
100 K, appears to be of the order of
$10^{20}$ cm$^{-2}$. This is  too low to be a major contributor to the
total X-ray absorption, though a small contribution is possible (see
\S 3.2).

 We also tried to fit the data with other models. Models with a black-body or
Raymond-Smith plasma at $z=0$ and Galactic column density fit badly
(Table~1). If N$_H$ is allowed to be free, it is found to be lower
than the Galactic for black-body model and higher than Galactic for
Raymond-Smith plasma, and the fits are good. Fluxes corresponding to
best fit model in each
case are given in Table~2. Note the extremely large temperature of the
optically thin thermal plasma at $z_l=0.886$. This, and the enormous
luminosity (1.4$\times 10^{45}$ erg s$^{-1}$) at $z_l=0.886$,
clearly exclude the possibility that the lensing galaxy is the X-ray
source (see Kim \etal~1992a,b and Fabbiano \etal~1992 for the X-ray
properties of normal galaxies).

\section{Discussion}

\subsection{Could the X-ray Source be Galactic?}

 The line of sight to PKS1830-211 passes close to the Galactic center.
The optical field in this direction is thus crowded, though the X-ray field
is not. Only one strong source is detected in the
2$^{\circ}$  PSPC field of view centered on PKS1830-211 and the
nearest faint source is $\sim 6^{\prime}$ away.  Even though
the positional coincidence of PKS1830-211 and the X-ray source is
excellent (\S 2), the spatial resolution of PSPC is only $\sim
25^{\prime \prime}$. It is thus possible that a Galactic source within
$25^{\prime \prime}$ from PKS1830-211 is  the X-ray source. Stars,
cataclysmic variables (CVs) both magnetic and non-magnetic, X-ray binaries,
are possible candidates. A detailed study of the field in optical and
infrared was reported by Djorgovski \etal~(1992). They find that the
brightest object in a 20 arcsec square field around PKS1830 is
$\sim$20 magnitude in V, identified to be a foreground M star. We note
that (1)~the X-ray source is non-flaring (2)~ the corresponding $\log
f_x/f_{opt}=$ 1.4 $-$ 1.6 (see Maccacaro et al. nomogram, 1988)
(3)~absorption in excess of the Galactic is indicated
for a Raymond-Smith optically thin thermal plasma. All these facts
together argue against the X-ray source being Galactic.

We conclude that the X-ray source is indeed PKS1830-211 and discuss
the implications in the following.

\subsection{The Source Quasar PKS1830-211}

 The X-ray luminosity of PKS1830-211 for various power-law models is given
in Table~2. The source is not detected in the optical with R magnitude
fainter than about 22$-$23 (Djorgovski et al. 1992). The V$-$R colors of
quasars in the atlas of Einstein sources have a broad distribution
with (V$-$R)$\sim0.4\pm0.2$ (Elvis \etal~ 1994). The V magnitude of
PKS1830-211 is then \gax 23.
Assuming that the source is at its optical detection limit, and the
X-ray luminosity given in Table~2, we can calculate the effective two
point optical to X-ray slope \aox~ using the formula in Wilkes \etal~ (1994).
If the optical extinction towards the source is only A$_V=2.7$
magnitudes (SNRS90) contributed by our Galaxy, the
resulting value of \aox~ is extremely small,  with \aox = 0.48 if the source
is at $z_s=1$ (0.46 for $z_s=1.5$). This is well outside the observed
range of \aox~ (Wilkes \etal~ 1994). There is, however, additional
extinction corresponding to the excess column density observed in the X-rays.
The effective column density at $z=0$ is about 10$^{22}$ cm$^{-2}$ (Table 1),
corresponding to total effective A$_V=5.8$\\ (E(B-V)=
max[0, (-0.055+1.987$\times10^{-22}N_H$)] and A$_V=3E(B-V)$).
The resulting \aox = 0.96 if the source is at $z_s=1$ (0.94 for $z_s=1.5$).
This is also very small, but within the observed distribution of \aox~
for radio loud quasars (Wilkes \etal~ 1994). The V$-$R color of the
Elvis \etal~ sample, however, may not represent a z$>1$ quasar. To
calculate the \aox~ directly from the observed R magnitude we modified
the formula in Wilkes \etal~ using the zero point R magnitude scale
given in Elvis \etal~ and extinction correction A$_R=2.32 E(B-V)$
(Savage \& Mathis 1979). The resulting
 \aox = 0.89 for $z_s=1$ (0.87 for $z_s=1.5$), again very small but
within the observed range.
Given that average \aox\ is 1.4 for a
radio loud quasar, PKS1830-211 is clearly over bright in X-rays unless the
absorbing galaxy has a gas to dust ratio  which differs from the average
Galactic value, yielding a larger optical extinction. PKS1830-211 is
one of the ten brightest sources in the sky at centimeter
wavelengths. So it is possible that it is intrinsically X-ray brighter
than an average radio-loud quasar.
Note that the above estimate of \aox~  assumes that the
optical magnitude of the source is at the plate limit. If the source
is in fact fainter, these numbers would be even more atypical.

Absorption and extinction are very important parameters in the
observed X-ray and optical fluxes of the source. It is interesting to
note that the z=0.886 equivalent hydrogen column density inferred from
the X-ray
data (N$_H= 3.5\times 10^{22}$ ~cm$^{-2}$) is very similar to the equivalent
H$_2$ column density inferred from the millimeter observations (WC96).
However, the molecular absorber of WC96
covers only one of the two images of the lensed quasar. The PSPC
observations cannot spatially resolve the two compact images. In the above
analysis, therefore, both the images were taken together as one X-ray source.
If the X-ray absorber covers only one image, the actual
column density of the X-ray absorber would be  larger than
we suggest above. To investigate this further, we fitted the X-ray
spectrum with two power-law models, one with and one without any
excess absorption along the line of sight. The photon index and the
normalization of the two power-laws were identical. The contour plot of
column density as a function of photon index of the absorbed power-law
is shown in Figure~2
(photon index $\Gamma$= 1+$\alpha_E$). It is evident that the fit
is not acceptable for a range of $\Gamma$ normal for  radio-loud
quasars. This implies either that (1)~PKS1830-211 has an extremely flat and
unusual spectrum, if the X-ray absorber covers only one of its images,
or that (2)~the X-ray absorber covers both the images.
In the latter case, the size of the absorber must be at least as
large as the separation between the two compact images, which is about
1$^{\prime\prime}$ (with $z_l=0.886$, this
corresponds to about 10 kpc). Thus the present observations are in accord
with its being a gas$-$ and dust$-$rich galaxy, with  radius  \gax 5 kpc.
Another possibility is that the X-ray absorption towards the NE image is
contributed by the z=0.19 absorber and that towards the SW image by
the z=0.886 absorber. The present X-ray data, however, cannot
constrain the two absorption components independently.

\section{Conclusions}

 We have shown that the X-ray source at the position of PKS1830-211 is
indeed the quasar. Its X-ray spectrum shows strong absorption, which at
the redshift $z_l= 0.886$ of the lensing galaxy corresponds to
N$_H=3.5^{+0.6}_{-0.5}\times 10^{22}$ atoms cm$^{-2}$ (for $\alpha_E=1.2$). The
corresponding $z=0$ effective optical extinction is large,
A$_V=5.8$. When corrected
for this additional extinction, the \aox\ of PKS1830-211 lies
within the observed range for quasars. We infer that both of the compact
images of PKS1830-211 must be covered by the X-ray absorber, the properties
of which are consistent with its being a late$-$type gas$-$ and dust$-$rich
galaxy. This is well in accord with conclusions from modeling (SNRS90, NNR93,
KN93) and radio observations (WC96). Present observations do not
constrain inclination of the lensing galaxy. They also do not rule out
the possibility of multiple lenses towards PKS1830-211.

The increasing numbers of gravitationally lensed systems with isolated
spiral galaxies as the lenses is a matter of considerable interest (Jackson
et al. 1997), as theoretical work on the statistics of lenses has tended so far
to neglect their role relative to that of early$-$type galaxies
(Turner \etal~1984, Kochanek 1996 and references therein).
As more and more `normal' late$-$type galaxies turn up as the
gravitational lenses, it is time to revise our understanding of the overall
mass distribution, and hence the lensing capabilities, of the average
spiral galaxy. PKS1830-211 is particularly interesting in that it appears
that there are two possible lenses along the line of sight to the source,
{\it both} of which appear to be rich in either gas or dust and possibly
disk systems. Future X-ray missions would allow us to understand this
system better. AXAF can resolve the two images. If the quasar has a
Fe-K emission line, it can be readily observed by ASCA, AXAF and
XMM. Observation of an emission line in X-rays might be our only hope
to determine the redshift of PKS1830-211.

\begin*{Acknowledgements:~}

 SM would like to thank her ``Galactic'' colleagues at CfA who
helped ruling out the possibility of the X-ray source being a Galactic
object.  S. Saar, E. Schlegel, P.  Callanan, B. Boroson, J.
Bookbinder, J. McClintock and J. Schachter provided useful discussions
on stars, CVs, and X-ray binaries.
M. Elvis and B. Wilkes are also thanked for their continual
encouragement and help. She gratefully acknowledges the financial
support of the NASA grant NAGW-4490 (LTSA).

 This research has made use of (1) data obtained through the High Energy
Astrophysics Science Archive Research Center Online Service, provided
by the NASA-Goddard Space Flight Center and (2) the NASA/IPAC
Extragalactic database (NED) which is operated by the Jet Propulsion
Laboratory, CALTECH, under contract with the National Aeronautics and
Space Administration.

\end*

\newpage

\noindent
{\bf Figure Captions:}\\

\noindent
{\bf Figure 1:} Data, best fit single power-law spectrum and residuals
to the fit. \\

\noindent
{\bf Figure 2:} Confidence contours for the two power-law fit (see the text).
 Contours of 68\%, 95\% and 99\% confidence regions are
shown.  \\

\newpage
\begin{table}[h]
\caption{Spectral fits to ROSAT data of PKS1830-211}
\begin{tabular}{|lcccc|}
\hline
model fitted&$\alpha_E~or~ kT^{a}$&$N_H(free)^{b}$&
$Norm.^{c}$&$\chi^2$(dof) \\
\hline
Power-Law:&&&&\\
+N$_H$(z=0, free)& 0.3$\pm1$ & 7$^{+0}_{-3}$ & 12$^{+19}_{-6}$ & 15.3(20) \\
+N$_H$(Gal, fixed)$^d$& -1.2$\pm0.3$ & & 4.1$^{+0.4}_{-0.5}$ & 20.3(21) \\
+N$_H$(Gal, fixed)$^d$& 0.2$\pm1$ & 23.5$^{+0.1}_{-19}$& 11$^{+20}_{-6}$ &
15.6(20)\\
{}~~~+N$_H$(z=1.0, free)&&&&\\
+N$_H$(Gal, fixed)$^d$& 0$^{+2}_{-1}$ & 17$^{+29}_{-15}$ & 10$^{+25}_{-19}$ &
16.2(20) \\
{}~~~+N$_H$(z=0.886, free)&&&&\\
+N$_H$(Gal, fixed)$^d$& 1.2 (Fixed)& 35$^{+6}_{-5}$&
21$\pm3$& 17.0(21)\\
{}~~~+N$_H$(z=0.886, free)&&&&\\
+N$_H$(Gal, fixed)$^d$& 0.2$\pm1$ & 7$^{+0}_{-5}$ &
 11$^{+15}_{-3}$ & 15.6(20)\\
{}~~~+N$_H$(z=0.19, free)&&&&\\
+N$_H$(Gal, fixed)$^d$& 1.2 (Fixed)& 11$\pm2$&
22$\pm3$& 17.4(21)\\
{}~~~+N$_H$(z=0.19, free)&&&&\\
+N$_H$(z=0, free)& 1.2 (Fixed)& 10$\pm$1 & 22$\pm$3 & 16.6(21)\\
&&&&\\
Black Body:&&&&\\
+N$_H$(z=0,free)& 3$^{+5}_{-1}$ & 1.3$\pm 1.0$& 11$\pm7$& 24.1(20) \\
+N$_H$(Gal,fixed)$^d$ & 3$\pm2$& & 12$^{+3}_{-7}$& 32.0(21)\\
&&&&\\
Optically Thin Thermal:&&&&\\
+N$_H$(z=0,free)& 0.8$\pm0.3$& 19$^{+4}_{-6}$& 106$\pm78$&
19.2(20) \\
+N$_H$(Gal,fixed)$^d$& 64$\pm 1$& & 3.4$\pm 0.2$& 102(21)\\
\hline
\end{tabular}
\smallskip

\small
\noindent
\newline
a. $\alpha_E$ is the energy spectral index ($f=Norm.E^{-\alpha_E}$);
\newline ~~temperatures are in keV.
\newline
b. $\times 10^{21}$ atoms cm$^{-2}$.
\newline
c. normalization units are: for power-law fits: 10$^{-4}$ keV cm$^{-2}$
s$^{-1}$ keV$^{-1}$ at 1~keV; \newline
{}~~for black body fits: 10$^{-4}$ keV cm$^{-2}$ s$^{-1}$ at kT keV.
\newline
{}~for optically thin thermal fits: 10$^{-17}$cm$^{-5}$
($(\int n_e^2 dV)/(4\pi D_L^2)$).
\newline
d. with N$_H$(Gal) fixed at 2.63$\times 10^{21}$ atoms cm$^{-2}$.
\newline

\end{table}

\newpage
\begin{table}[h]
\caption{PKS1830-211 X-ray Fluxes and Luminosities}
\begin{tabular}{|lccc|}
\hline
Model & f$_x$(0.2--2.4 keV)$^a$ & L$_x$(0.2--2.4 keV)$^b$ & L$_x$(0.2--2.4
keV)$^b$ \\
& & z=1.0 & z=1.5 \\
\hline
PL (N$_H$ free, $\alpha$ free) & 1.4 & 2.3 & 6.1 \\
PL$^c$ ($\alpha$ free)& 1.4& 1.6 & 4.1 \\
PL$^c$ ($\alpha=1.2$ fixed)& 1.26& 10.2 & 32.8\\
PL (N$_H$ free, $\alpha=1.2$ fixed) & 1.26 & 10.2 & 32.8\\
\hline
Raymond-Smith$^d$ (z=0)& 1.16 & 3.7 &\\
Raymond-Smith (z=0.886)& 1.34& 0.14 & \\
Blackbody$^d$ (z=0)& 1.58& 0.23 & \\
&& &\\
\hline
\end{tabular}
\smallskip

\small
\noindent
a. $10^{-12}$ erg cm$^{-2}$ s$^{-1}$, observed.
\newline
b. $10^{46}$ erg s$^{-1}$, corrected  for absorption.
\newline
c. Power-law model with Galactic absorption fixed, plus additional
absorption at z=0.886.
\newline
d. Luminosity corresponding to a distance of 10 kpc in units of
$10^{34}$ erg s$^{-1}$.
\end{table}

\newpage
%\vspace*{5in}
\begin{figure}
\centerline{
\psfig{figure=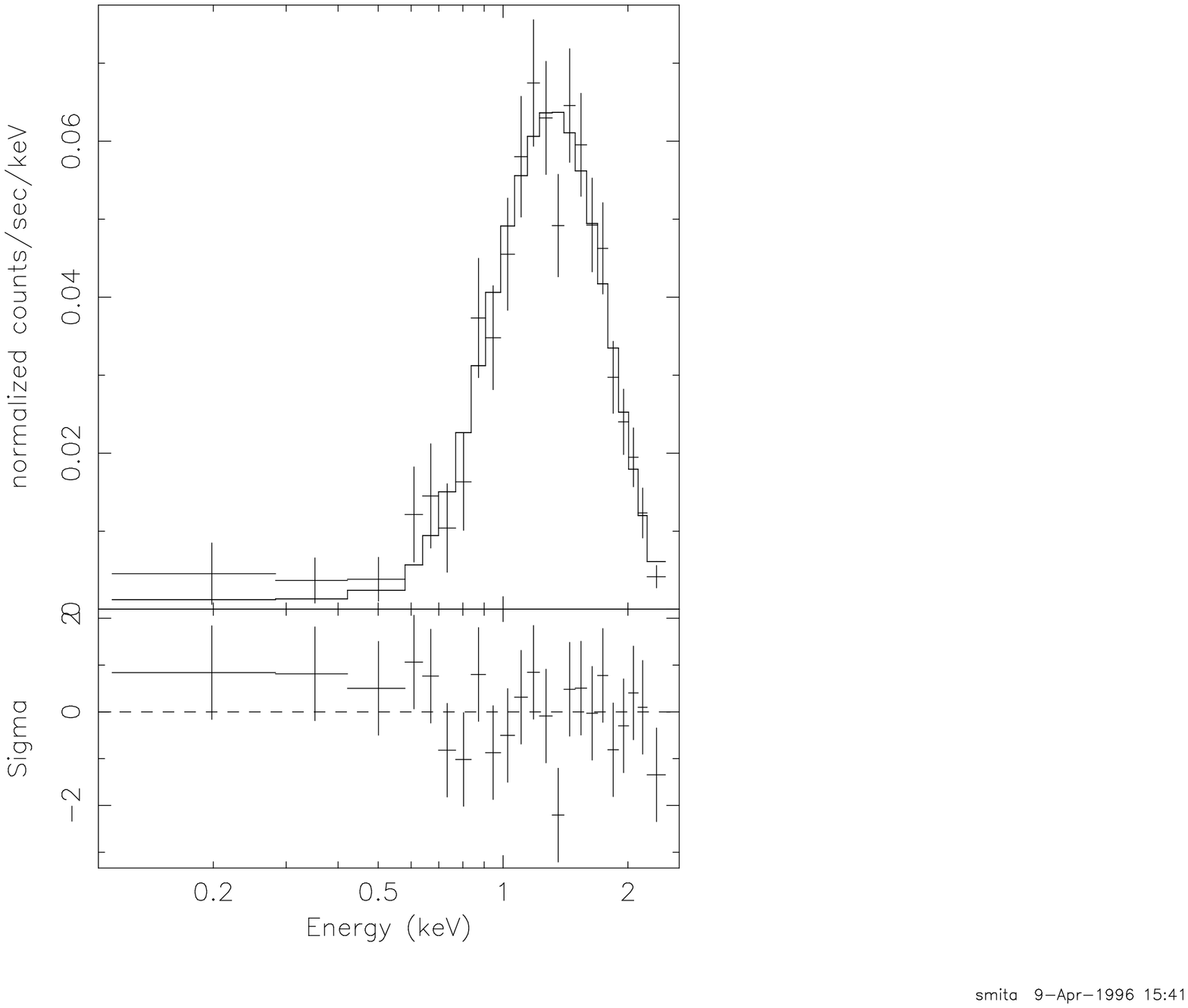,height=6.5truein,width=4.25truein,angle=-90}
}
\caption{ }
\end{figure}

\newpage
\begin{figure}
\centerline{
\psfig{figure=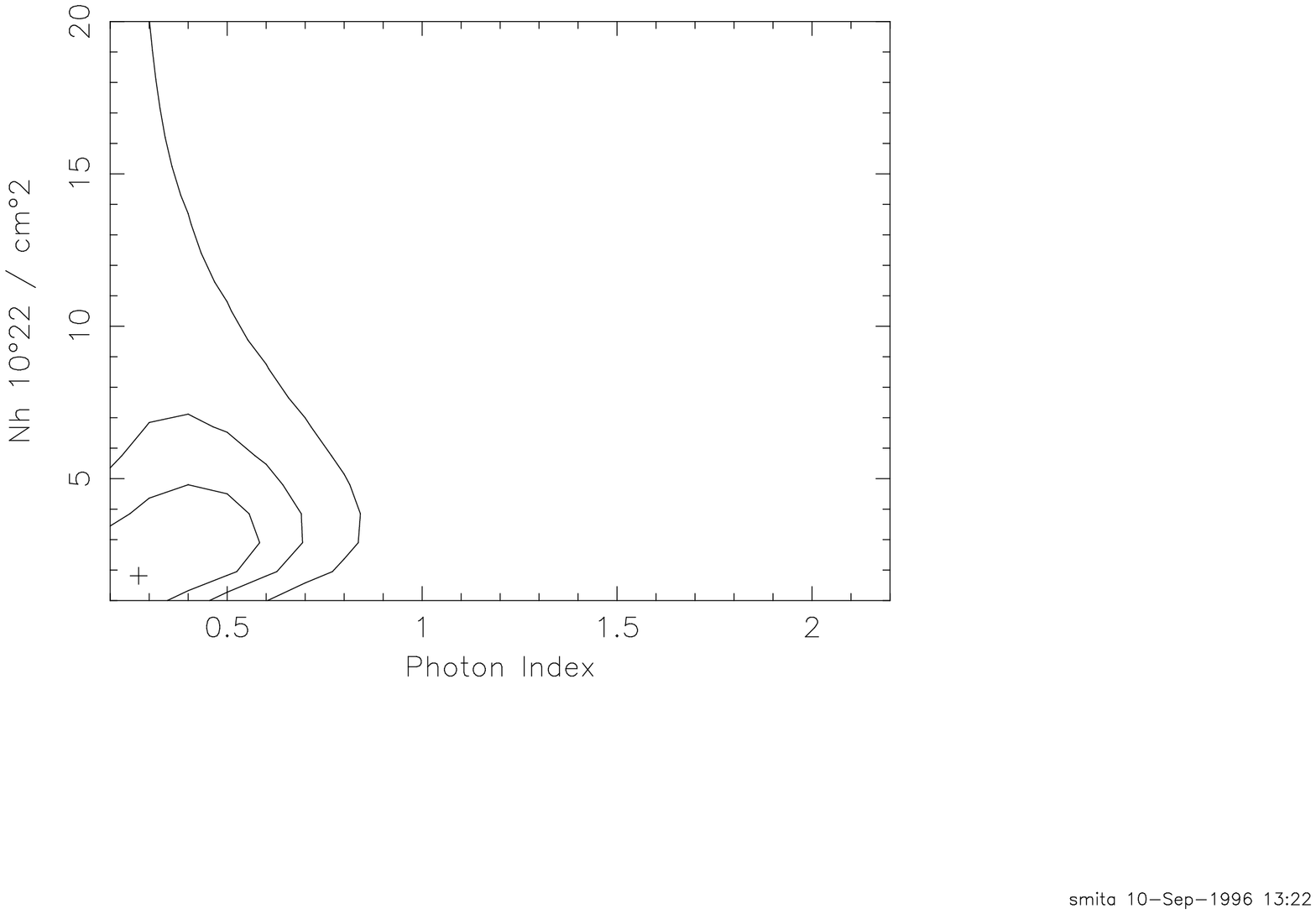,height=6truein,width=6truein,angle=-90}
}
\caption{ }
\end{figure}


\newpage
\begin{references}

\reference{}Brinkmann, W., Siebert, J., \& Boller, Th. 1994, A\&A, 281, 355
\reference{}Djorgovski, S. et al. 1992, MNRAS, 257, 240
\reference{}Elvis,~M., Wilkes,~B.~J., McDowell,~J.~C., Green,~R.~F.,
Bechtold,~J., Willner,~S.~P., Cutri,~R., Oey,~M,~S., \& Polomski,~E.
{}~1994, ApJS, 95, 1
\reference{}Fabbiano, G., Kim, D-W., \& Trinchieri, G. 1992, ApJS, 80, 531
\reference{}Jackson, N. et al. 1997, ApJL, $submitted$
\reference{}Jauncey, D. L. et al. 1991, Nature, 352, 132
\reference{}Kim, D-W., Fabbiano, G., \& Trinchieri, G. 1992a, ApJ, 393, 134
\reference{}Kim, D-W., Fabbiano, G., and Trinchieri, G. 1992b, ApJS, 80, 645
\reference{}Kochanek, C. \& Narayan, R. 1993, ApJ. 401, 461 (KN93)
\reference{}Kochanek, C.S. 1996, preprint, CfA
\reference{}Lovell, J. E. J. \etal~1996, ApJL, in press
\reference{}Morrison, R. \& McCammon, D. 1983, ApJ, 270, 119
\reference{}Nair, S., Narasimha, D. \& Rao P. 1993, Ap. J. 407, 46 (NNR93)
\reference{}van Ommen et al. 1995, ApJ, 444, 561
\reference{}Pfeffermann,~E. \etal~ 1987 Proc. SPIE Int. Soc. Eng. 733, 519
\reference{}Rao, A.~P. \& Subrahmanyan, R. 1988, MNRAS, 231, 229
\reference{}Refsdal, S. 1964, MNRAS, 128, 307.
\reference{}Refsdal, S. 1966, MNRAS, 132, 101.
\reference{}Savage, B. D. \& Mathis, J. S. 1979, ARA\&A, 17, 73
\reference{}Schartel, N. \etal~1996, MNRAS, in press
\reference{}Shastri P., Wilkes, B., Elvis, M. \& McDowell, J. ~1993,
ApJ, 410, 29
\reference{}Stark, A. A. \etal~1992, ApJS, 79, 77
\reference{}Subrahmanyan, R., Narasimha, D., Rao, P., \& Swarup, G.
1990, MNRAS, 246, 263 (SNRS90)
\reference{}Tr\"umper, J. 1983, Adv. Space Res., 2, No. 4, 241
\reference{}Turner E.L., Ostriker J.P. \& Gott J.R., 1984, ApJ, 284, 1
\reference{}Wiklind, T. \& Combes, F. 1996, Nature, 379, 139 (WC96)
\reference{}Wilkes,~B.~J., Tananbaum,~H., Worrall,~D.~M., Avni,~Y.,
Oey,~M.~S. \& Flanagan,~J. 1994 ApJS,~ 92, 53

\end{references}
\end{document}